\documentclass[10pt]{iopart}
\usepackage{epsfig}
\usepackage{graphicx}
\usepackage{dcolumn}
\usepackage{bm}
\usepackage{color} 

\usepackage{ulem}

\usepackage[switch]{lineno}
\begin{document}

\title[]{Spin structure and spin Hall magnetoresistance of epitaxial thin films of the insulating non-collinear antiferromagnet SmFeO$_3$}

\author{T Hajiri$^1$, L. Baldrati$^2$, R. Lebrun$^2$, M. Filianina$^{2,3}$, A. Ross$^{2,3}$, N. Tanahashi$^1$, M. Kuroda$^1$, W. L. Gan$^4$, T. O. Mente\c{s}$^5$, F. Genuzio$^5$, A. Locatelli$^5$, H Asano$^1$, M. Kl\"{a}ui$^{2,3,6}$}

\address{$^1$ Department of Materials Physics, Nagoya University, Nagoya 464-8603, Japan}
\address{$^2$ Institut f\"{u}r Physik, Johannes Gutenberg-Universit\"{a}t Mainz, Staudingerweg 7, Mainz, D-55128, Germany}
\address{$^3$ Graduate School of Excellence Materials Science in Mainz, D-55128 Mainz, Germany}
\address{$^4$ School of Physical and Mathematical Sciences, Nanyang Technological University, 21 Nanyang, 637371, Singapore}
\address{$^5$ Elettra Sincrotrone Trieste, I-34149 Basovizza (TS), Italy}
\address{$^6$ Center for Quantum Spintronics, Department of Physics, Norwegian University of Science and Technology, 7034 Trondheim, Norway}
\ead{t.hajiri@nagoya-u.jp}
\ead{klaeui@uni-mainz.de}
\vspace{10pt}
\begin{indented}
\item[]January 2019
\end{indented}

\begin{abstract}
We report a combined study of imaging the antiferromagnetic (AFM) spin structure and measuring the spin Hall magnetoresistance (SMR) in epitaxial thin films of the insulating non-collinear antiferromagnet SmFeO$_3$. 
X-ray magnetic linear dichroism photoemission electron microscopy measurements reveal that the AFM spins of the SmFeO$_3$(110) align in the plane of the film. 
Angularly dependent magnetoresistance measurements show that SmFeO$_3$/Ta bilayers exhibit a positive SMR, in contrast to the negative SMR expected in previously studied collinear AFMs. 
The SMR amplitude increases linearly with increasing external magnetic field at higher magnetic field, suggesting that field-induced canting of the AFM spins plays an important role. 
In contrast, around the coercive field, no detectable SMR signal is observed, indicating that SMR of AFM and canting magnetization components cancel out. 
Below 50~K, the SMR amplitude increases sizably by a factor of two as compared to room temperature, which likely correlates with the long-range ordering of the Sm ions. 
Our results show that the SMR is a sensitive technique for non-equilibrium spin system of non-collinear AFM systems.
\end{abstract}

%
\vspace{2pc}
\submitto{\JPCM}
%
%
\ioptwocol

\section{Introduction}
Antiferromagnetic (AFM) spintronics is an emergent field of research, aiming to exploit AFMs in magnetic memory and logic applications, thanks to the advantageous magnetic properties of AFMs over ferromagnets (FMs) such as the absence of stray magnetic fields, terahertz (THz) spin dynamics, and low magnetization switching currents~\cite{AFM_spintronics1, AFM_spintronics2, AFM_spintronics3, Romain_Nature}. 
While the vanishing net magnetization makes it hard to control and detect AFM magnetic moments electrically, recent experiments have demonstrated the electrical control of AFMs by spin-orbit torques~\cite{CuMnAs_SOT, Mn2Au_SOT, NiO_SOT, Moriyama_SOT} and the electrical detection of the N$\rm{\acute{e}}$el vector orientation by spin-Hall magnetoresistance (SMR) ~\cite{NiO_SOT, Moriyama_SOT} and anisotropic magnetoresistance~\cite{TAMR_Park, SIO_Fina, PRB_Hajiri}.
These results highlight the potential of AFMs to replace FMs in spintronic applications.

In FM/heavy metal (HM) bilayers, the SMR is well known to depend on the relative angle between the FM magnetization and the spin polarization direction $\mu_s$ of the spin accumulation from spin Hall effect (SHE) in the HM layer~\cite{YIG_SMR1, YIG_SMR2}. 
Since the FM magnetization aligns parallel to the external magnetic field and the SHE generates a spin current ${\bm j}_s \propto {\bm j}_c \times {\bm \mu}_s$, the SMR exhibits a cos$^2\theta$ angular variation, where $\theta$ is the angle between charge current $j_c$ and external magnetic field $\mu_0H$. 
In the case of AFMs, the magnetic moments are aligned mostly along the N$\rm{\acute{e}}$el vector, which tends to align perpendicular to the external magnetic field. 
Therefore, a phase shift of 90$^\circ$ with respect to the FM case, i.e. a sin$^{2}⁡\theta$ variation as a function of the angle, has been predicted~\cite{Negative_theory}, which is termed negative SMR. 
Up to now, there are several experimental SMR studies in AFM/HM bilayers.
The negative SMR has been demonstrated in different NiO samples~\cite{Negative_NiO, multidomain_NiO, Lorenzo_NiO}, while both positive and negative SMR have been reported in SrMnO$_3$~\cite{Positive_SrMnO3}, FeMn~\cite{FeMn_SMR} and Cr$_2$O$_3$~\cite{Positive_Cr2O3, Negative_Cr2O3}.
A negative SMR was also reported in the ferrimagnetic insulator gadolinium iron garnet (GdIG) around the compensation temperature, while a positive SMR was observed far from the compensation temperature~\cite{Bowen_GdIG, canted_ferri_SMR}.
At the compensation temperature, GdIG exhibits a non-collinear canted magnetization. 
Although these results indicate that the sign of the SMR is sensitive to the spin configurations, it is not clear why some AFMs exhibit a positive SMR. 
On the other hand, AFM SMR studies have been performed only on collinear AFM materials.
Since the weak FM moments resulting from canted AFM spins could theoretically lead to a positive SMR~\cite{Romain_Nature}, the SMR study on AFM with weak FM moments resulting from canted AFM spins may enable us to clarify it.

The orthorhombic rare-earth orthoferrite SmFeO$_3$ (SFO) is an AFM insulator with N$\rm{\acute{e}}$el temperature $T_{\rm N}=670$~K~\cite{SFO_Neel}. 
The canted AFM ordering of Fe spins through an inverse Dzyaloshinskii-Moriya interaction (DMI) causes improper ferroelectricity at the same temperature as the AFM ordering~\cite{SFO_PRL}.
Other orthorhombic rare-earth orthoferrites exhibit ferroelectricity only at low temperature (e.g., GdFeO$_3$ at $T=2$~K~\cite{GdFeO3_TC}, DyFeO$_3$ under magnetic field at $T~\sim4$~K~\cite{DyFeO3_TC}), although they have a high N$\rm{\acute{e}}$el temperature $T_{\rm N}=661$~K and 645~K, respectively. 
Thus, SFO is one of the most interesting materials among rare-earth orthoferrites. 
SFO exhibits a weak FM moment due to four nonequivalent Fe spins. 
In the bulk, at temperatures below 140~K, a long-range ordering of nonequivalent Sm spins appears~\cite{SFO_PRL, SFO_long-range, SFO_spin}. 
Up to now, the spin structure of SFO films and the relation between external magnetic fields and the resulting spin structure that leads to the SMR signal have not been discussed yet. 
Thus, the effect of the interplay between AFM and the long-range ordering of Sm spins as well as a weak FM on SMR calls for temperature dependent studies of this material.

In this work, we show that SFO(110) thin films exhibit an AFM easy plane anisotropy by using X-ray magnetic linear dichroism photoemission electron microscopy (XMLD-PEEM) imaging of the spin structure. 
In the AFM easy plane of SFO, a positive SMR signal of SFO/Ta bilayers is obtained in a temperature range of 50--300 K, indicating that the weak FM due to canted AFM spins results in a positive SMR above the monodomainization field.
On the other hand, below the monodomainization field, no detectable SMR signal is observed.
The SMR amplitude linearly increases with external magnetic fields up to 10~T, showing  clear difference with conventional AFMs like NiO. 
Although a weak FM magnetization slightly increases with decreasing temperature, the SMR amplitude is constant above 150~K and increases by a factor 2 at 50~K, which indicates an effect of long-range ordering of Sm ions at low temperature.

\section{EXPERIMENTAL DETAILS}
High-quality epitaxial SFO films were prepared on LaAlO$_3$ (LAO) (001) substrates by magnetron sputtering using a polycrystalline SFO target. 
More details on the sample preparation can be found in Ref.~\cite{SFO_growth}. 
After performing the film growth, we deposited in-situ a Pt capping layer (5~nm) for XMLD-PEEM measurements and a Ta layer (3~nm) for SMR by magnetron sputtering at room temperature.
The Pt capping layer was used to prevent charging effects during XMLD-PEEM measurements and was thinned by Ar ion etching before the measurements. 
The crystal structure was analyzed using both in-plane and out-of-plane X-ray diffraction (XRD) measurements using Cu~$K\alpha$ radiation. 
XMLD-PEEM measurements were performed using as-grown bilayers at the Nanospectroscopy beamline of the Elettra synchrotron light source equipped with an Elmitec PEEM setup (type LEEM III)~\cite{Elettra1, Elettra2}. 
The magnetic properties of the SFO/Ta bilayers were characterized using superconducting quantum interference device (SQUID) magnetometry. 
The SMR measurements were performed in a variable temperature insert cryostat with a superconducting magnet. 
The sample was rotated in the strong magnetic field by means of a piezoelectric rotator. Before the SMR measurements, the bilayers were patterned into Hall bars (550~$\mu$m $\times$ 75~$\mu$m) by electron beam lithography and ion beam etching.

\section{RESULTS AND DISCUSSIONS}

\begin{figure}[t]
\begin{center}
\includegraphics[width=\linewidth]{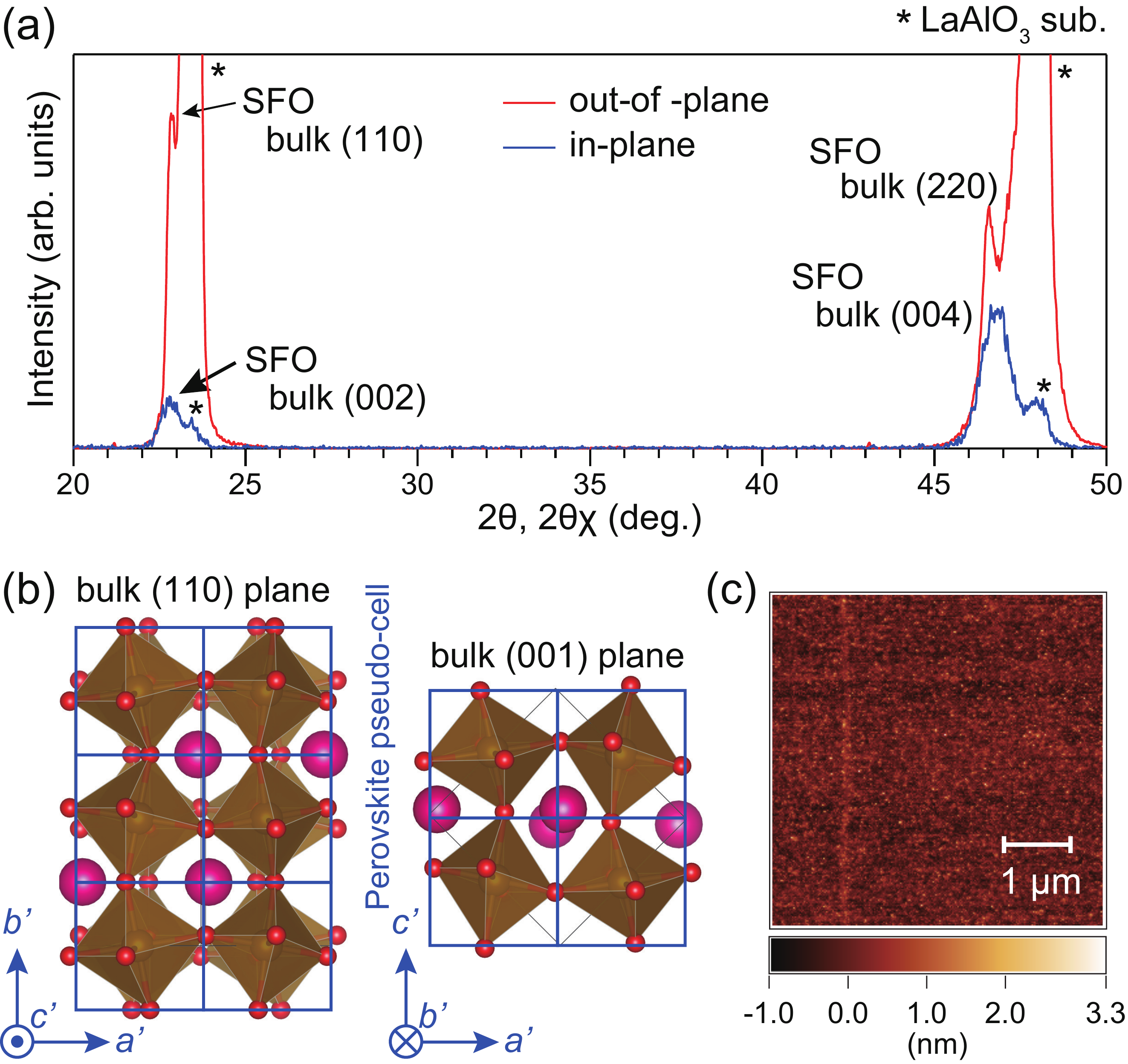}
\caption{
(a) Out-of-plane and in-plane XRD profiles of 55~nm thick SFO thin films. 
(b) Crystal structure of SFO viewed in bulk $\langle110\rangle$ and $\langle001\rangle$ directions, visualized using the software package VESTA~\cite{VESTA}. 
Bold solid squares indicate perovskite pseudo-cells. 
(c) Surface morphology of 150~nm thick SFO films.
}
\label{fig:one}
\end{center}
\end{figure}

Figure~\ref{fig:one}(a) shows the out-of-plane and in-plane XRD patterns of a 55~nm thick SFO film on a LAO substrate. 
Only the bulk (110) and (001) SFO peak series show Bragg peaks in the out-of-plane and in-plane XRD patterns with $d_{220}=0.1946$~nm and $d_{004}=0.1933$~nm on LAO (001) substrates, respectively. 
From the XRD profiles, we determine the epitaxial relationship to be LAO substrate (001) [100] $\parallel$ bulk SFO (110) [001]. 
Figure~\ref{fig:one}(b) shows the crystal structures of the SFO viewed in the bulk $\langle110\rangle$ and $\langle001\rangle$ directions. 
Bulk (110)-oriented SFO corresponds to the (001)-oriented perovskite pseudo-cell ($a^{\prime}$, $b^{\prime}$, $c^{\prime}$), as shown by the bold solid lines. 
The lattice match between perovskite pseudo-cell and LAO substrates enables us to obtain high quality epitaxial films. 
Hereafter, we denote by $a^{\prime}$, $b^{\prime}$ and $c^{\prime}$ the crystal orientation of the perovskite pseudo-cell to distinguish it from the bulk crystal orientation.
Figure~\ref{fig:one}(c) shows the surface morphology of the SFO layer (150~nm) measured by using an atomic force microscope. 
The root mean square value of the roughness is estimated to be 0.356~nm, comparable to other SMR studies of AFM insulators~\cite{Negative_NiO, Negative_Cr2O3}.

\begin{figure}[t]
\begin{center}
\includegraphics[width=\linewidth]{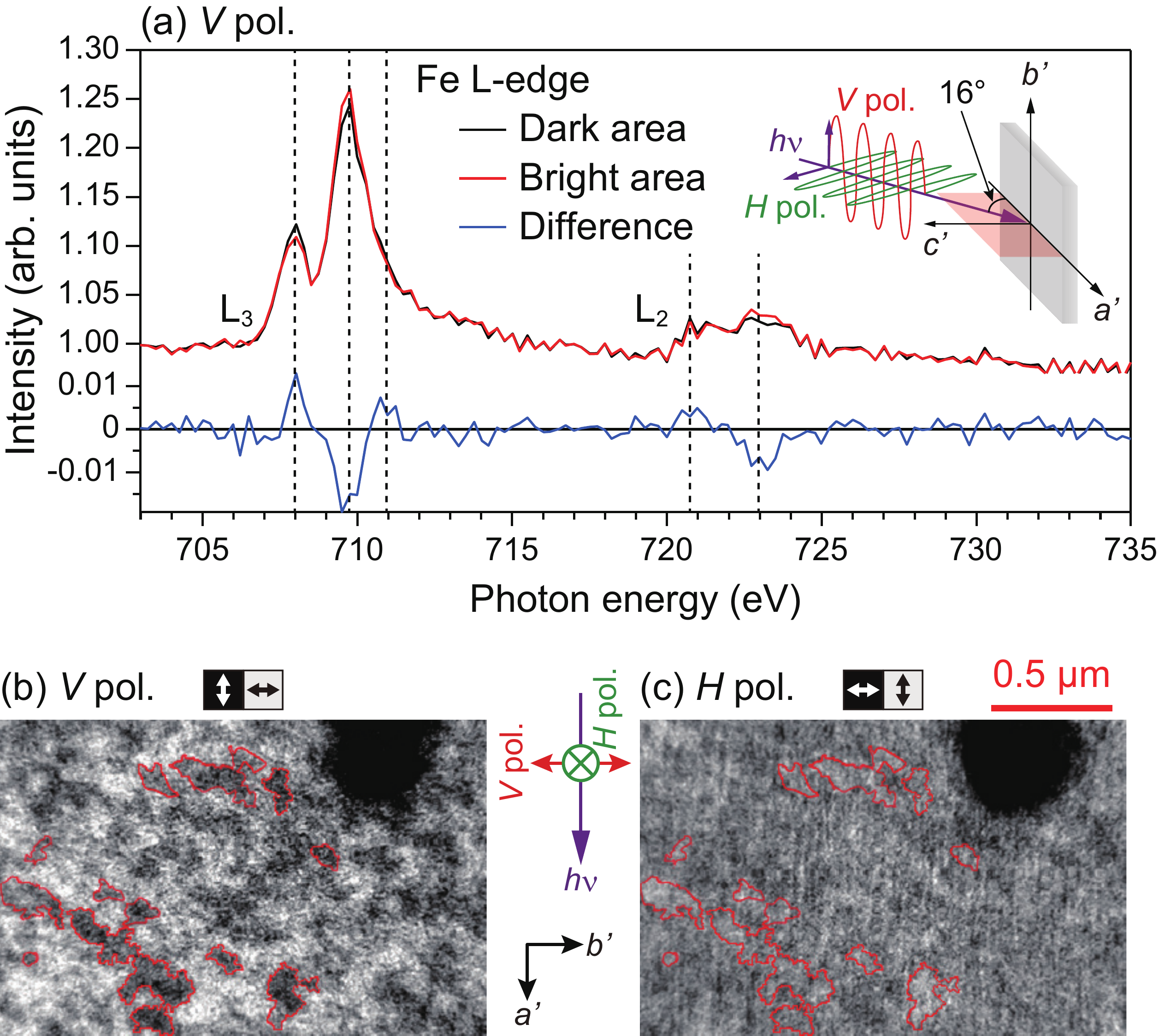}
\caption{
(a) XAS and XMLD spectra for SFO/Pt films at the Fe-$L_{2,3}$ edge taken by vertically polarized x-rays. 
The inset shows the experimental configuration of the XMLD-PEEM system and the coordination of perovskite pseudo-cell. 
(b,c) XMLD-PEEM images of SFO/Pt films taken by (b) vertical and (c) horizontal polarizations with $h\nu_1=708.0$~eV and $h\nu_2=709.8$~eV (see text for details). 
The dark spot at the top right of the images is a mark for comparison between linear vertical and horizontal polarizations.
The white (black) arrows in the black (gray) boxes indicate the N$\rm{\acute{e}}$el vector orientation
All measurements were performed on as-grown bilayers.
}
\label{fig:two}
\end{center}
\end{figure}

Bulk SFO exhibits G-type AFM order at 300~K, where the Fe spins align antiferromagnetically along the bulk $c$ direction. 
Since SFO grows as bulk (110)-oriented films on LAO(001) substrates, the AFM spins are expected to align parallel to $a^{\prime}$ and $b^{\prime}$ in the film plane.
However, there are only a few studies of SFO films and no reports on AFM spin structure of SFO films.
To confirm the AFM spin orientation, XMLD-PEEM measurements were performed on SFO(90 nm)/Pt(5 nm) samples. 
Figure~\ref{fig:two}(a) presents X-ray absorption spectra at the Fe~$L$ edge, acquired by secondary electron emission using linear vertical polarization. 
The multiplet splitting of the Fe~$L_3$ and $L_2$ edges is clearly visible. 
By calculating the asymmetry of two images taken at the same linear polarization and two different photon energies corresponding to the multiplet splitting, XMLD-PEEM images can be obtained, according to Ref.~\cite{LFO_domain, Mn2Au_XMLD};
\begin{eqnarray}
{\rm XMLD}=(I_{h\nu_2}-I_{h\nu_1})/(I_{h\nu_1}+I_{h\nu_2}).
\label{eq:one}
\end{eqnarray}
From the XAS spectra as shown in Fig.~\ref{fig:two}(a), the photon energies $h\nu_1$ and $h\nu_2$ to obtain the maximum XMLD contrast with respect to the multiplet splitting of Fe~$L_3$ edge are determined to be 708.0~eV and 709.8~eV, respectively. 
The XMLD images acquired with linear vertical and horizontal polarizations are shown in Figs.~\ref{fig:two}(b) and \ref{fig:two}(c), respectively. 
Here, the electric field component of the vertical polarization is parallel to the $b^{\prime}$ direction while the horizontal polarization has both $a^{\prime}$ and $c^{\prime}$ components, as shown in the inset of Fig.~\ref{fig:two}(a). 
In the asymmetry image acquired with vertical polarization as a difference between the two photon energies noted above, domains with an average size of about 100~nm are clearly visible.
On the other hand, the contrast is much weaker when the images are acquired with linear horizontal polarization. 
The XAS spectra extracted from bright and dark areas of Fig.~\ref{fig:two}(b) as well as their difference are presented in Fig.~\ref{fig:two}(a). 
The relationship of the XAS spectra between the dark and bright areas is the same as for typical XMLD spectra, acquired with electric field of the X-ray beam alternatively parallel or perpendicular to the AFM axis. 
In addition, the size of the contrast areas seen in the XMLD images is similar to the AFM domains of the orthorhombic rare-earth orthoferrite LaFeO$_3$~\cite{LFO_domain}. 
This allows us to conclude that the obtained asymmetry contrast regions in linear vertical polarization originate from the AFM domains.

\begin{figure}[t]
\begin{center}
\includegraphics[width=\linewidth]{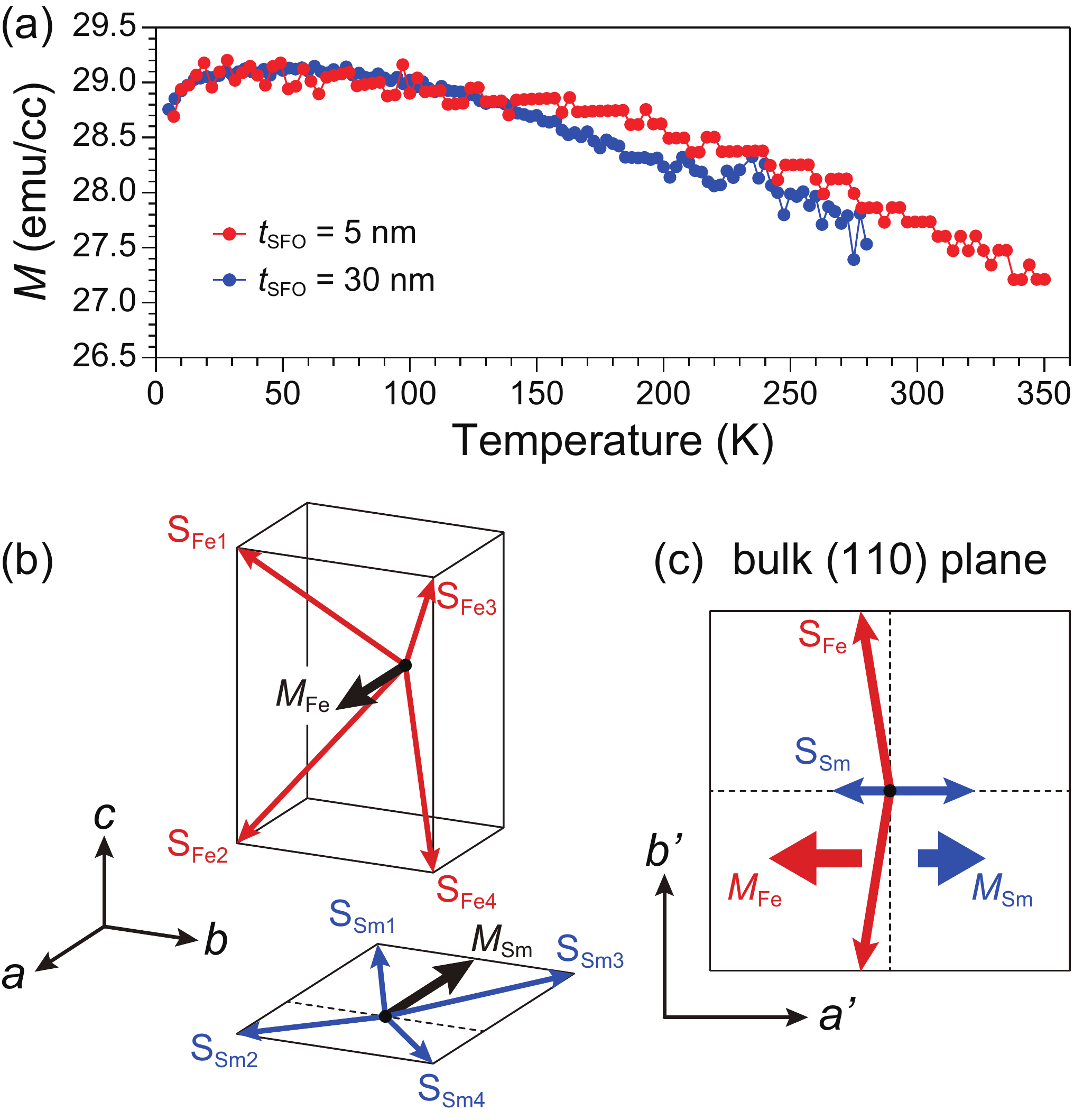}
\caption{
(a) Temperature dependence of the magnetization of SFO/Ta bilayers with an external magnetic field of 10~mT parallel to the $a^{\prime}$ direction. 
The magnetization of $t_{\rm SFO} =30$~nm is normalized to the one of $t_{\rm SFO} =5$~nm for comparison.
(b) Schematic spin structure of bulk SFO~\cite{SFO_PRL}.
(c) Proposed spin structure of thin film SFO viewed in the $\langle110\rangle$ direction.
}
\label{fig:three}
\end{center}
\end{figure}

As shown in the inset of Fig.~\ref{fig:two}(a), the X-ray beam is incident at a grazing angle of 16$^\circ$ with respect to the sample surface. 
While the linear vertical polarization has an electric field component only in the plane of the sample, the linear horizontal polarization has both an out-of-plane and a small in-plane electric field component. 
Therefore, the weak XMLD contrast in the linear horizontal polarization can be considered to originate from the in-plane AFM component of the magnetic moments only.
This is confirmed by the inverted contrast between Fig.~\ref{fig:two}(b) and Fig.~\ref{fig:two}(c), as can be seen in the domains surrounded by lines, because the electric field components parallel to the sample surface of the vertical and horizontal polarizations are parallel to the $b^{\prime}$ and $a^{\prime}$ directions, respectively. 
These results highlight that the AFM magnetic moments of the SFO are aligned in the $a^{\prime}b^{\prime}$ easy magnetic plane, as expected from the bulk spin configuration~\cite{SFO_PRL, SFO_long-range}. 
Note that no ferromagnetic contrast is observed in X-ray magnetic circular dichroism (XMCD) PEEM images.

Figure~\ref{fig:three}(a) shows the temperature-dependent magnetization of SFO(5 and 30~nm)/Ta(3~nm) bilayers. 
Both films exhibit a similar temperature dependence; the magnetization monotonically increases with decreasing temperature and starts to slightly decrease at around 65~K. 
In the bulk, the long-range ordering of nonequivalent Sm spins appears at temperatures below 140 K~\cite{SFO_PRL, SFO_long-range}. 
This leads to a reduction of the total magnetization at low temperature, since, as shown in Fig.~\ref{fig:three}(b), the directions of weak FM components due to nonequivalent Fe spins ($M_{\rm Fe}$) and Sm spins ($M_{\rm Sm}$) are opposite. 
Therefore, the observed decrease of the magnetization in SFO/Ta bilayers indicates that the long-range ordering of the Sm$^{3+}$-spin moments exists in the SFO thin films at least below 65~K as well. 
The maximum magnetization of SFO(5~nm)/Ta(3~nm) bilayers is approximately 29~emu/cc, which is 4~times larger than in the bulk. 
The value of the magnetization observed in thin films is larger than that of the bulk material, which has been attributed to the compressive strain applied by the LAO substrate~\cite{SFO_growth}. 
Based on this higher magnetization value and considering the 8.2~mrad canting angle of bulk SFO~\cite{canting_angle1, canting_angle2}, we can estimate the canting angle in the SFO(5~nm)/Ta(3~nm) bilayers to be 34.0~mrad. 
From the XMLD-PEEM imaging and the temperature-dependent magnetization of the SFO thin films, we estimate a spin structure of SFO films as schematically shown in Fig.~\ref{fig:three}(c), which is consistent with the bulk spin structure~\cite{SFO_PRL, SFO_long-range}. 
We note that the four nonequivalent Sm spins are considered to be located in the $ab$ plane in the bulk~\cite{SFO_PRL, SFO_long-range, SFO_spin}. 
Although it is not clear whether four nonequivalent Sm spins are located in the $a^{\prime}b^{\prime}$ plane like in the bulk in our SFO film, we only considered Sm spins projected onto the $a^{\prime}b^{\prime}$ plane.

\begin{figure}[t]
\begin{center}
\includegraphics[width=\linewidth]{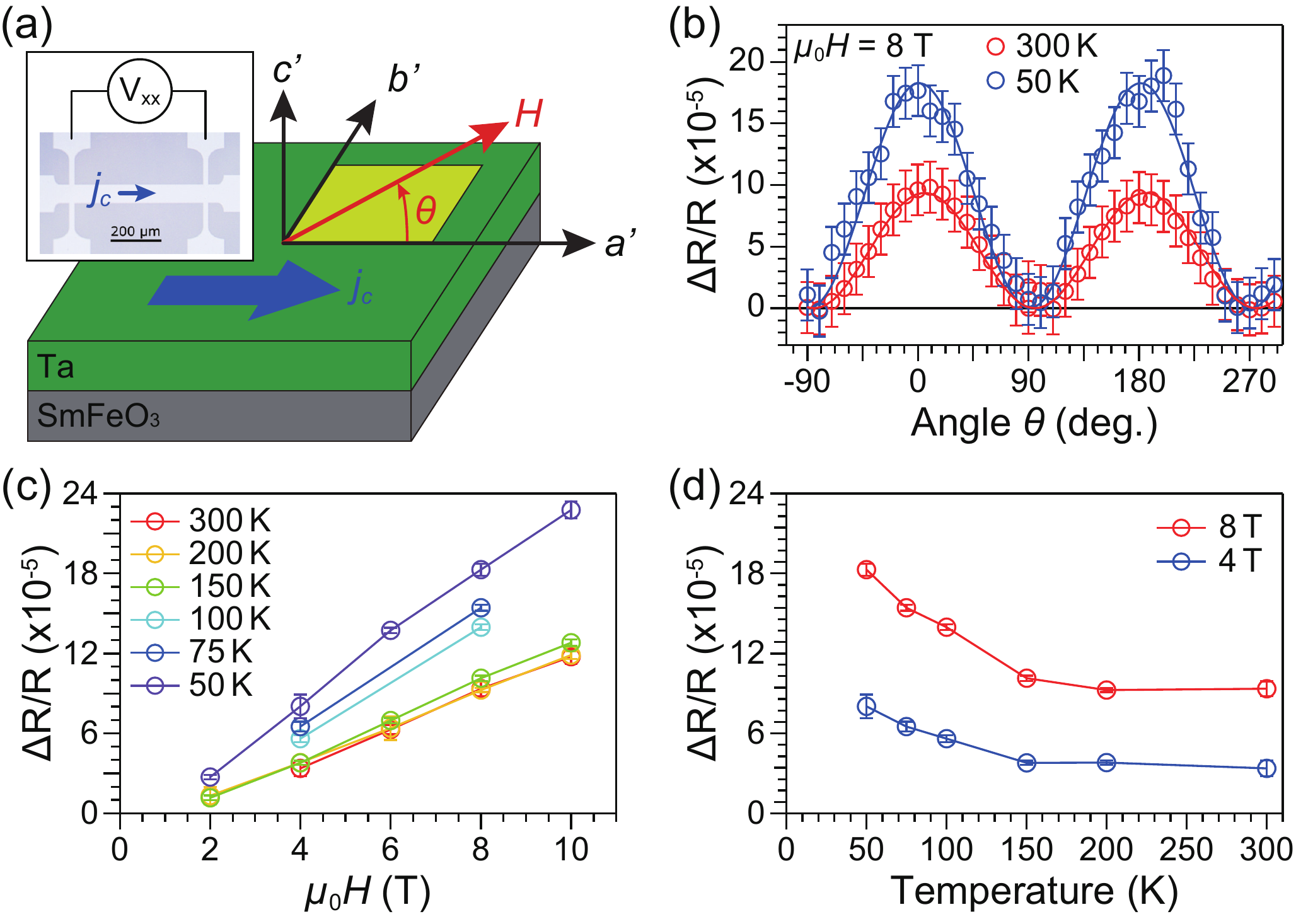}
\caption{
(a) Experimental configuration of SMR measurements. 
The inset shows the optical microscope image of the Hall bar design.
(b) Angular dependences of the SMR amplitude of SFO(5~nm)/Ta(3~nm) bilayers at a field of 8~T, at 300~K and 50~K. 
The solid lines are fitting curves using the equation $\Delta R/R=A{\rm sin}^2\theta$.
(c) Field dependence of the SMR amplitude obtained from ADMR. 
(d) Temperature dependence of the SMR amplitude at 4~and 8~T.
}
\label{fig:four}
\end{center}
\end{figure}

A key requirement to using this material for applications is the ability of electrically detecting the magnetic order. Here we probe the competing SMR signal from the AFM N$\rm{\acute{e}}$el order parameter and the weak FM moment. To explore this, we performed angularly dependent magnetoresistance (ADMR) measurements of SFO(5~nm)/Ta(3~nm) bilayers. 
The experimental configuration is shown in Fig.~\ref{fig:four}(a). 
The resistivity of the SFO(5~nm)/Ta(3~nm) bilayers increases with decreasing temperature and its value (378~$\mu\Omega$cm at 300~K and 410~$\mu\Omega$cm at 50~K) is in agreement with that of $\beta$-phase Ta~\cite{Negative_Cr2O3}, indicating that the Ta film is in the $\beta$-phase. 
It is well known that $\beta$-phase Ta has one of the largest spin Hall angles and the strongest spin-orbit coupling among all the possible Ta phases. 
Figure~\ref{fig:four}(b) shows the angular dependence of the ADMR amplitudes in the presence of an external magnetic field $\mu_0H=8$~T. 
Note that Ta has been shown not to exhibit a significant anisotropic magnetoresistance arising from the magnetic proximity effect in YIG/Ta bilayers above 10~K~\cite{YIG/Ta_MPE}. 
In the same way, it is reported that the influence of magnetic proximity effect can be excluded in Cr$_2$O$_3$/Ta bilayers~\cite{AHE_Cr2O3}. 
Although finite ADMR was observed in $\beta$-phase Ta films on SiO$_2$ substrates under $\mu_0H=9$~T~\cite{Ta_Hanle}, its magnitude is negligibly small ($\Delta R/R=4.5 \times 10^{-6}$ for $t_{\rm Ta}=5$~nm). 
It is also known that the particular morphology of the film may have profound influence on the magnetoresistive properties~\cite{Ta_Hanle}, but we are not aware of reports on Ta with a pronounced ADMR.
We thus attribute the origin of our signal to the spin Hall magnetoresistance (SMR). 
The open circles in Fig.~\ref{fig:four}(b) are experimental data and the solid lines are fitting curves of equation $\Delta R/R=A{\rm cos}^2\theta$. 
We find maximum and minimum resistances when the external magnetic field is parallel or perpendicular to the charge current $j_c$, respectively. 
The angular dependence of the ADMR amplitudes can be fitted well by a cos$^2⁡\theta$ dependence, indicating a positive SMR behavior, which is expected in FMs, in contrast to the negative SMR, varying with the in-plane angle as sin$^2\theta$, previously found in collinear AFMs/HM bilayers~\cite{Negative_NiO, multidomain_NiO, Lorenzo_NiO, Negative_Cr2O3}. 

The magnetic field dependence of the SMR amplitude obtained from ADMR is shown in Fig.~\ref{fig:four}(c). 
The SMR amplitude linearly increases with the applied magnetic field up to 10~T at all the temperatures used in this study. 
According to the multidomain AFM SMR theory based on the redistribution of magnetostrictive domains~\cite{multidomain_NiO, Lorenzo_NiO}, the SMR amplitudes in AFMs are expected to increase quadratically with the applied magnetic field. 
This behavior is due to the quadratic dependence of the Zeeman energy in AFMs, having no net magnetic moment in zero field, and was confirmed by experiments in single crystal and thin film NiO~\cite{Negative_NiO, multidomain_NiO, Lorenzo_NiO}. 
The Zeeman energy in a material like SFO may additionally lead to a linear field-dependent term, stemming from the field-induced canting of the magnetic moments, which does not lead to a measurable SMR signal in NiO. 
On the other hand, the complicated domain structures due to the competition between AFM and weak FM should appear below the monodomainization field~\cite{Helen_theory}, which is discussed later.

\begin{figure}[t]
\begin{center}
\includegraphics[width=\linewidth]{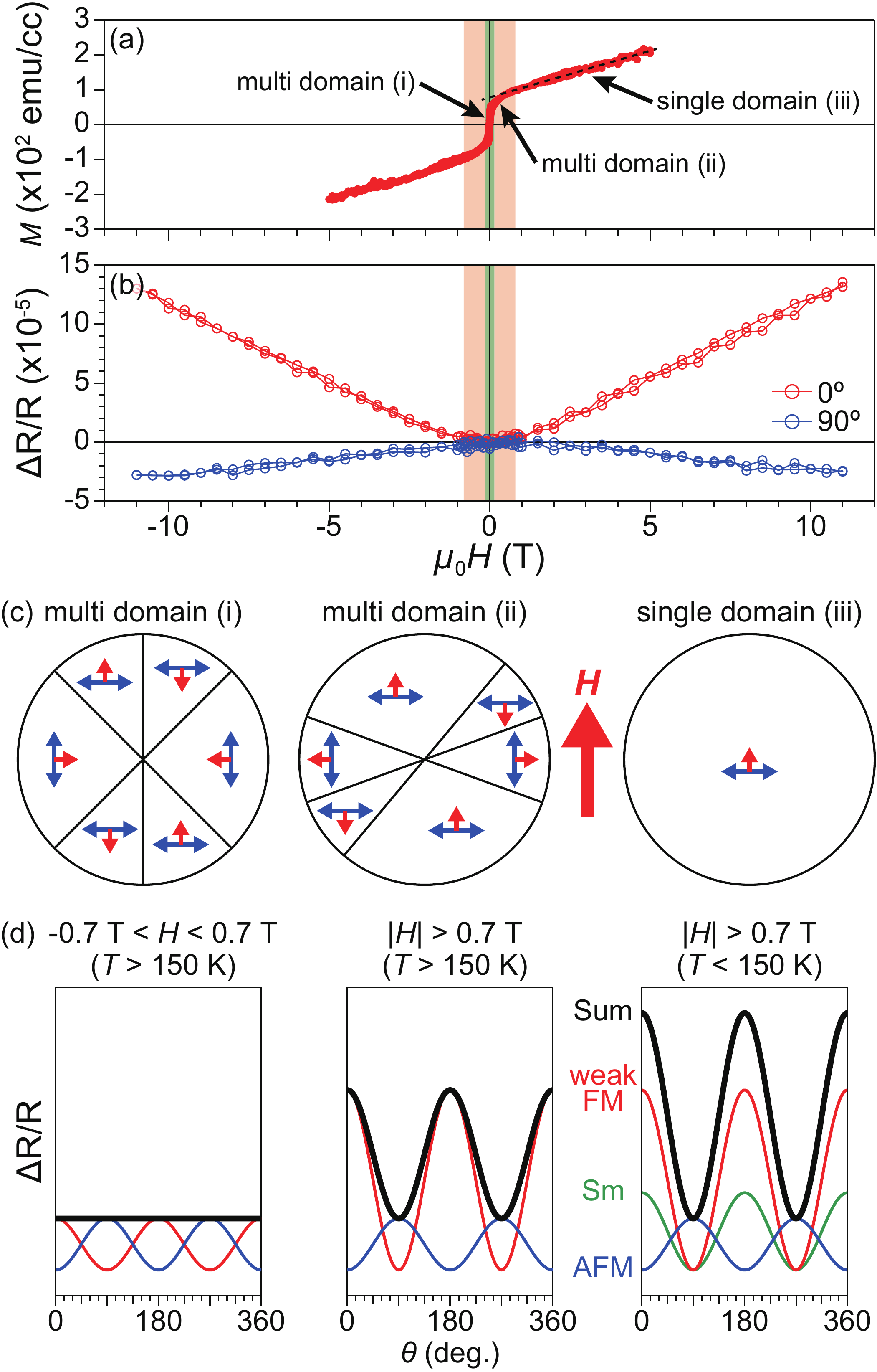}
\caption{
(a) Hysteresis loop of $t_{\rm SFO} = 5$~nm measured at $T=200$~K.
(b) Uniaxial magnetoresistance field scan parallel and perpendicular to the current.
(c) Behavior of the competed domain structures between AFM and weak FM in the external magnetic field~\cite{Helen_theory}.
(d) Angular dependence of the expected SMR amplitude components from AFM, weak FM and Sm and sum of them.
}
\label{fig:five}
\end{center}
\end{figure}

In addition to the linear field dependence of the SMR amplitude, an unusual temperature dependence of the SMR signal is also observed. 
Figure~\ref{fig:four}(d) presents the temperature dependence of the SMR amplitude at two different fields. 
While the SMR amplitude is relatively independent of temperature above 150~K, the SMR amplitude increases below 100~K and is almost two times larger at 50~K.
The origin of the temperature dependence of the SMR is still debated, since both the HM and the insulator play a role~\cite{Tem_depe_SMR1, Tem_depe_SMR2}. 
In the case of NiO/Pt bilayers, the negative SMR amplitude increases linearly with decreasing temperature, without any sudden increase versus temperature~\cite{Negative_NiO}. 
Recently, it was reported that Cr$_2$O$_3$/Ta bilayers exhibit a large temperature dependence from positive SMR at 300~K to negative SMR below 250~K, which is attributed to the competition between the surface FM and bulk AFM moment dependence on the temperature of Cr$_2$O$_3$~\cite{Negative_Cr2O3}. 
In contrast to NiO and Cr$_2$O$_3$, bulk SFO exhibits a weak FM due to canting of the Fe AFM moments. 
In addition, long-range ordering of the Sm$^{3+}$-spin moments resulting from four nonequivalent Sm spins appears below 140~K in bulk SFO~\cite{SFO_PRL, SFO_long-range, SFO_spin}. 
As discussed in Fig.~\ref{fig:three}, signatures of long-range ordering of the Sm$^{3+}$-spin moments are observed in our SFO/Ta bilayers at least below 65~K. 
Since the SMR amplitude is constant at temperatures above 150~K while a weak FM magnetization slightly increases with decreasing temperature, the increasing weak FM magnetization plays no major role in determining the SMR amplitude. 
Moreover, there are no significant changes in SMR amplitudes below 225~K in Cr$_2$O$_3$/Ta bilayers~\cite{Negative_Cr2O3}, suggesting that Ta plays no role in SMR amplitudes at 50~K.
On the other hand, as discussed in Fig.~\ref{fig:three}, the magnetization of Sm$^{3+}$ is aligned parallel to Fe canted magnetic moment but opposite direction.
This configuration can cause the increment of the positive SMR amplitude.
These results indicate that the increasing positive SMR amplitudes are related to the long-range ordering of the Sm ions rather than the increasing weak FM magnetization. 

Finally, we would like to discuss the possible origin of positive SMR amplitudes in SFO/Ta bilayers.
Figure~\ref{fig:five}(a) and \ref{fig:five}(b) show the hysteresis loop along $a^{\prime}$ direction and the magnetic field dependence of the resistance curve along $\theta=0^\circ$ and $90^\circ$ at $T=200$~K, respectively.
The hysteresis loop shows linear dependence above 0.7~T, suggesting the domain structure becomes monodomain.
Below 0.7~T, the gradients become steeper.
The coercive field is about 0.15~T.
In the magnetic field dependence of the resistance curve, no sizable change of SMR signal is observed below 0.7~T and the linear dependence of the resistance is observed above 0.7~T. 
According to the theoretical consideration of domain redistribution of the competition between AFM and weak FM~\cite{Helen_theory}, there are four-type domains and the domain redistribution behavior in external magnetic field is expected as shown in Fig.~\ref{fig:five}(c).
Since a previous AFM SMR study pointed out that the origin of AFM SMR is magnetic-field-induced AFM domain redistribution, a sizable SMR amplitude is expected for $|H|<0.7$~T in our SMR results.

The negative SMR was predicted by a theoretical study~\cite{Negative_theory} and demonstrated by experimental studies of NiO single crystals~\cite{Negative_NiO} and epitaxial films~\cite{multidomain_NiO}, and stems from the fact that the localized AFM spins in an easy plane align perpendicularly to the applied magnetic field, in contrast to the magnetic moment of a ferromagnet, which aligns parallel to the field. 
SFO films exhibit an AFM easy plane in the $a^{\prime}b^{\prime}$ plane as deduced from our XMLD-PEEM imaging, which is expected to cause a negative SMR.
When the canting-induced FM magnetization exists in AFM, the interfacial spin current $j_s$ between AFM and HM can be theoretically written as the sum of AFM and FM components:
\begin{eqnarray}
\fl
j_s=\frac{g_n}{4\pi}{\bm n\times(\bm\mu \times \bm n)}+\frac{g_{rm}}{4\pi}{\bm m\times(\bm\mu \times \bm m)}+\frac{g_{im}}{4\pi}{\bm\mu\times m}\nonumber\\
&&
\label{eq:two}
\end{eqnarray}
where ${\bm n}$ is the N$\rm \acute{e}$el vector, ${\bm m}$ the magnetization due to the canted magnetic moment, $g_n$ the N$\rm \acute{e}$el spin-mixing conductance, $g_{rm}$ and $g_{im}$ the real and imaginary parts of the magnetic spin-mixing conductance~\cite{Romain_Nature}.
$g_n$ is associated to show the negative SMR, while $g_{rm}$ and $g_{im}$ describe the positive SMR.
If AFM and FM components have almost the same SMR amplitudes, the AFM and FM contributions can cancel out as shown in left-panel of Fig.~\ref{fig:five}(d), so that no sizable change of the SMR signal is observed below 0.7~T.
Then, above 0.7~T, the FM component becomes larger than the AFM one, which results in the positive SMR as shown in middle-panel of Fig.~\ref{fig:five}(d).
Therefore, the field induced canting is likely to explain the linear dependence of the SMR amplitudes. 
The monodomainization field does not depend on the temperature.
So the effect of the long-range ordering of Sm ions below 100~K can be understood by simply adding the positive SMR component as shown in right-panel of Fig.~\ref{fig:five}(d). 
However, SMR signals should saturate at the monodomainization fields from previous reports~\cite{Negative_NiO, multidomain_NiO}. 
From our data, we can conclude that 10 T are not enough to saturate the field-induced canting, which is also not expected due to the strong AFM exchange.
The absence of XMCD contrast in PEEM measurements implies that the presence of FM surface states can be excluded. 
Compared to hematite~\cite{Romain_Nature, Romain_arXiv} and GIG~\cite{Bowen_GdIG, canted_ferri_SMR}, where the FM-like SMR is negligible compared to the AFM-like one, our results suggest that the FM-related torques in SFO dissipate the spin current more efficiently than the AFM-related ones, resulting in the positive SMR we observe. 
This difference is possibly related to the different interfacial coupling mechanism between the conduction electrons in the heavy metal and the magnetic order of the insulator, which involves several components and can depend on the materials used, as currently being debated ~\cite{Cheng_PRL 2014, Kamra_PRL 2017}.

\section{CONCLUSION}
We performed a combined study of spin structure imaging and SMR measurements in epitaxial SFO(110) films, where the Fe spins are ordered antiferromagnetically with a weak FM magnetization, resulting from DMI.
We determine by XMLD-PEEM that the Fe magnetic moments are oriented in an easy plane parallel to the surface. 
In the AFM easy plane of the SFO, a positive SMR signal of the SFO/Ta bilayers is obtained. 
The SMR amplitude increases linearly with the external magnetic fields up to 10~T above 0.7~T due to field-induced canted Fe magnetic moment, which is different from the quadratic dependence found in AFMs such as NiO. 
On the other hand, no detectable SMR signal is observed below 0.7~T, where the competing domains redistribute, indicating that the AFM and weak FM components of the SMR cancel out.
We find that the SMR amplitude is constant above 150~K up to room temperature, but becomes two times larger when decreasing the temperature down to 50~K. 
This increase of the SMR amplitude at low temperatures may correlate with the long-range ordering of the Sm ions, indicating that the rare earth ions increase the SMR values in rare earth orthoferrites. 
Our results clearly show that the positive and negative SMR are both present in the material and compete due to their different symmetry. We thus demonstrate that the SMR is an useful technique to explore the long-range ordering as well as canting weak magnetic moments. 
This opens the way to further SMR studies in this class of materials.

\section{ACKNOWLEDGMENTS}
The authors thank Prof. H. Gomonay for fruitful discussions on the SMR results.
Part of this work was carried out at the Nanospectroscopy beamline of the Elettra synchrotron light source, Basovizza (TS). 
This work was supported by the Japan Society for the Promotion of Science (JSPS) Program for Advancing Strategic International Networks to Accelerate the Circulation of Talented Researchers, JSPS KAKENHI Grant Number 17K17801, the Murata Science Foundation, the Deutsche Forschungsgemeinschaft (DFG) SPP 1538 “Spin Caloric Transport” and SFB TRR 173 “Spin + X”, the Graduate School of Excellence Materials Science in Mainz (GSC 266), and the EU project INSPIN (FP7-ICT-2013-X 612759). 
L.B. and R.L. acknowledge the EU Marie Sk{\l}odowska-Curie grant agreement ARTES No. 793159 and FAST No. 752195, respectively.

\end{document}